\documentclass[preprint,12pt,3p]{elsarticle}

\usepackage{amssymb}

\journal{Physics Letters B}

\begin{document}

\begin{frontmatter}

\title{Is interacting vacuum viable? }

\author{Georgia Kittou}
\address{Department of Mathematics, College of Engineering, \\ American University of the Middle East\\ P.O.Box: 220 Dasman, 15423 Kuwait}
\date{today}

\ead{georgia.kittou@aum.edu.kw}

\begin{abstract}
We study the asymptotic dynamics  of dark energy as  a mixture of presssureless matter and an interacting vacuum component.  We find that the only dynamics compatible with  current observational data favors an asymptotically vanishing matter-vacuum energy interaction in a model where dark energy is simulated by a generalized Chaplygin gas cosmology.
\end{abstract}


\end{frontmatter}

\section{Introduction}
\label{sec1}
\par\noindent
In recent years most of the cosmological studies have been  focused  on variations of General Relativity and modifications of the Standard Cosmological model \cite{mod}. This is done in  order to provide a more reliable framework   to explain the present  physical  evidence of the   universe. It is observed today that only $5\%$ of the matter content is of baryonic form  and additional evidence coming from the high redshift surveys of type I supernovae \cite{obs1,obs2} indicate that we currently live in a universe that undergoes an accelerated expansion.
\par
In the present literature many  cosmological models involve the presence of an exotic type of matter component that lies beyond the framework of standard cosmology, cf. \cite{c1,c2,c3,c4}, in an attempt to explain the present acceleration of the observed universe. Additionally, the case of  coupling and energy transfer  between dark energy and dark matter leads to research efforts that try to alleviate the so-called coincidence problem  \cite{co1,co2}. One unified dark energy model that has attracted the interest for research is the  Generalized Chaplygin Gas model (GCG in short). This model has a dual character since at early times it satisfies the properties of a matter-dominated universe whereas at late times it approaches the  limiting behaviour of dark-energy dominated universe \cite{Cha}.
\par
In previous works \cite{p1,p2,phd}, we have studied the asymptotic dynamics near finite-time singularities of flat and curved universes filled with two interacting fluids using an interaction term that was first introduced by Barrow and Clifton, cf. \cite{ba1,ba2}.   In the present work, we consider the case of energy exchange between dark matter (as pressureless dust) and dark energy (as vacuum) with the local energy transfer being associated with the energy density of the vacuum (that is $\rho_V$) so that $Q_{\mu}=-\nabla_{\mu}\rho_V$ \cite{w1}.
\par
In the limit of zero energy exchange ($Q_{\mu}=0$), or equivalently if the vacuum energy is covariantly conserved ($\nabla_{\mu}\rho_V=0$), then the vacuum energy must be homogeneous in spacetime and equal to a cosmological constant \cite{Cha2}. Under these conditions, we address the question of the viability and stability of the interacting vacuum model on approach to the finite-time singularity by studying the asymptotic properties of solutions of the scale factor, the total energy density and total pressure of the universe. 
\par
The cosmological model is expected to be stable, and therefore acceptable, if  asymptotically it reproduces the dominant features of dark matter and dark energy at both early and late times respectively. We show that asymptotically at early times the energy exchange is vanishing and the energy density of the vacuum is approximately zero, in contrast to what occurs in the standard cosmological model. Hence, at early times and in the absence interaction,  our model is  indistinguishable from the CDM model \cite{Bento}.
\par
The asymptotic analysis of the solutions is carried out using the method of asymptotic splittings, cf. \cite{met1, met2}.
The analysis provides a complete description of all possible dominant features that the solution possesses as it is driven to a blow-up.
\par
The plan of this paper is as follows. In the next section, we write down all possible asymptotic decompositions of the basic differential equations of our problem describing the GCG model. Sections \ref{subsection1}-\ref{subsection3} present a detailed study of the various asymptotic solutions. In the last section we discuss our results and point out some interesting open problems in this field.

\section{Decomposed Dark energy models}
\label{sec2}
\par\noindent
We study the case of the generalised Chaplygin gas model  in flat FRW universe as a mean to explain the accelerated expansion of the universe \cite{2}. In the GCG approach the exotic cosmological fluid is
defined by the barotropic equation of state 
\begin{equation}
    P_{cgc}=A\rho_{cgc}^{-\alpha},
\end{equation}\par\noindent
where $A$ is a positive constant and $0<\alpha\leq1$. This leads to a cosmological solution for the density 
\begin{equation}\label{gcg}
    \rho_{cgc}=\left(A+Ba^{-3(\alpha+1)}\right)^{1/(\alpha+1)},
\end{equation}\par\noindent
where $a$ is the scale factor of the universe and $B$ is a positive integration of constant  for a well defined $\rho_{cgc}$ at all times. 
From Eq. (\ref{gcg}), one can conclude that at early times the asymptotic solution for the energy density  reproduces the CDM model as described by
\begin{equation}
    \rho_{cgc}\sim a^{-3}\quad a\rightarrow 0,
\end{equation}\par\noindent
in the limit of vanishing constant $\alpha\rightarrow0$ \cite{Sand,Park}. At late times  the solution (\ref{gcg})   implies that the fluid behaves as a cosmological constant
\begin{equation}
    \rho_{cgc}\sim A^{1/(\alpha+1)}\quad a\rightarrow\infty.
\end{equation}\par\noindent
This interpolation of the  model  between two different  fluids at different stages of the evolution of the universe suggests that the GCG model can be interpreted as a mixture of two cosmological fluids  with energy exchange.
\par
Now, any unphysical  oscillations or exponential blow-up in the matter spectrum produced by such a unified model \cite{Sand} can be avoided, if  one excludes  coupling with  phantom fields \cite{Bento}. Therefore, the unique coupling between dark matter (pressureless dust) and dark energy (cosmological constant) makes   the GCG model a well-behaved model both at early times (approach the successful CDM model) and at late times (approach de Sitter Universe).
\par
It is interesting to mention here that the interaction between the fluid components allows energy to  be transferred from dark matter to dark energy, since $\alpha$ is a positive constant.  As we will show below, this energy transfer is vanishingly small  at early times making the model indistinguishable from a CDM model in the past. Whereas when interaction starts off the contribution of the cosmological constant is significant and the model approaches de Sitter universe \cite{Bento}.
\par
The Einstein equations\footnote{Here we consider the case where the baryons have a similar behaviour to that of a pressureless dust, i.e dark matter and we exclude the possibility of an energy exchange between baryons and dark energy (see \cite{Bento} for more information.} for a flat Friedman universe filled with pressureless dust ($\rho_m$) and vacuum ($\rho_V$), scale factor $a(t)$ and Hubble expansion rate $H=\dot{a}/a$  reduce to the Friedman equation
\begin{equation}\label{f1}
    3H^2=\rho_m+\rho_V=\rho_{gcg}.
\end{equation}\par\noindent
The total energy momentum tensor of the pair is the algebraic some of the individual energy-momentum tensor given by
\begin{equation}
    T_{total}^{\mu\nu}= T_{m}^{\mu\nu}+ T_{V}^{\mu\nu}.
\end{equation}\par\noindent
Since the two fluids are not separately conserved it occurs that 
\begin{equation}
    \nabla_{\nu}T^{\mu\nu}_m=-u^{\mu}\quad \nabla_{\nu}T^{\mu\nu}_V=u^{\mu},
\end{equation}\par\noindent
where $u^{\mu}$ is the total 4-velocity so that $\nabla_{\nu}T^{\mu\nu}_{total}=0$.
\par 
It is shown in \cite{phd} that by taking the covariant derivative of each energy density momentum separately one obtains
\begin{equation}
    -u^0=\nabla_{\nu}T^{0\nu}_m=-a^3\dot{p_m}+\frac{d}{dt}[a^3(p_m+\rho_m)]
\end{equation}
and similarly for the second one
\begin{equation}
    u^0=\nabla_{\nu}T^{0\nu}_V=-a^3\dot{p_V}+\frac{d}{dt}[a^3(p_V+\rho_V)].
\end{equation}\par\noindent
The  forms of the continuity equations then read
\begin{eqnarray*}
    \dot{\rho_m}+3H\rho_m&=&-\frac{u^0}{a^3}\\
    \dot{\rho_V}&=&\frac{u^0}{a^3}.
\end{eqnarray*}\par\noindent
If we set $Q=u^0/a^3$ as the interaction term, one can show after some calculations that the interaction term reads
\begin{equation}\label{interaction function}
    Q=\frac{u^0 H}{\dot{a}a^2}.
\end{equation}\par\noindent
Therefore, the interaction function (\ref{interaction function}) is generally dependent on the expansion rate
$H$, the scale factor $a(t)$, its time derivative $dot{a(t)}$, as well as on the energy
densities  and pressures of the fluid components. We note here that if the expansion of the universe ceases, that is for $H=0$, the interaction between the fluid components will also vanish due to the fact that interaction is  coupled to the $3$-geometry of the slice with a  mean curvature described from the Hubble parameter $H$.
\par
We  assume a fluid interaction of the form \cite{w1}
\begin{equation}
Q=3\alpha H\left(\frac{\rho_m\rho_V}{\rho_m+\rho_V}\right),
\end{equation}\par\noindent
and the final forms of the continuity equations for matter and vacuum are given by
\begin{eqnarray}
    \dot{\rho_m}+3H\rho_m&=&-Q\label{f2}\label{c1}\\
    \dot{\rho_V}&=&Q\label{f3}\label{c2},
\end{eqnarray}\par\noindent
respectively. Equations (\ref{f1}),(\ref{f2}) and (\ref{f3}) describe a $3$-dimensional system with unknowns $(a, \rho_m, \rho_V)$ satisfying the constraint given by Eq. ($\ref{f1}$). After some manipulations it is proved that the above set of equations  leads to the following master differential equation
\begin{equation}\label{master}
    \ddot{H}+3(\alpha+1)H\dot{H}+2\alpha\frac{\dot{H}^2}{H}=0.
\end{equation}\par\noindent
Equation (\ref{master}) is a nonlinear differential equation of second order for the Hubble parameter $H$. If we assume a power-law type solution for the scale factor  $a=t^p$ that is $H=p/t$, where $p\in\mathcal{Q}$, then one can  provide all possible exact solutions for the Hubble parameter as well for the scale factor based on the GCG parameter $\alpha$ at both early and late times.
\par 
Indeed if we substitute the form  $H=p/t$ in Eq. (\ref{master}) we get the following equation for the exponent $p$
\begin{equation}
    p^2[2-3p(\alpha+1)+2\alpha]=0.
\end{equation}\par\noindent
We find two possible solutions to the equation above; The first one describes the case of no interaction with $p=0$ while the non-trivial solution satisfies the form
\begin{equation}
    p=\frac{2+2\alpha}{3(\alpha+1)},
\end{equation}\par\noindent
and the exact solution for the scale factor reads
\begin{equation}\label{scalef}
    a(t)=t^{(2+2\alpha)/[3(\alpha+1)]}.
\end{equation}\par\noindent
However, in this work we are interested in an asymptotic analysis of solutions of (\ref{master}) near finite-time singularities. To do so, it will be very useful for our calculations to rewrite the master equation (\ref{master}) in a suitable dynamical system form. In this respect, we rename $H=x$ and find the $2$-dimensional system 
\begin{eqnarray}
    \dot{x}&=&y\nonumber\\
    \dot{y}&=&-3(\alpha+1)xy-2\alpha\frac{y^2}{x}\label{ds}.
\end{eqnarray}\par\noindent
Equivalently, we have the vector field
\begin{equation}
    f(x,y)=[y,-3(\alpha+1)xy-2\alpha\frac{y^2}{x}]\label{vf}.
\end{equation}\par\noindent
The vector field can split \cite{met1}
in three different ways namely
\begin{eqnarray}
    \label{dc1}f_{I}(x,y)&=&[y,-3(\alpha+1)xy]+(0,-2\alpha\frac{y^2}{x}),\\
    \label{dc2}f_{II}(x,y)&=&(y,-2\alpha\frac{y^2}{x})+[0,-3(\alpha+1)xy],\\
    \label{dc3}f_{III}(x,y)&=&[y,3(\alpha+1)-2\alpha\frac{y^2}{x}].
    \end{eqnarray}
  \par\noindent  
In the following sections we apply the method of asymptotic splittings, analytically expounded in \cite{met1, met2}, to describe the asymptotic properties of the solutions of the dynamical system (\ref{ds}) in the vicinity of its finite-time singularities.
\section{Early times asymptotics}\label{subsection1}     
\par\noindent
In this section, we give necessary conditions in terms of the parameter $\alpha$ for the existence  of generalised Fuchsian series  type solutions \cite{met2} towards the finite-time singularity of the first decomposition  $f_{I}(x,y)=[y,-3(\alpha+1)xy]+(0,-2\alpha\frac{y^2}{x})$. 
\par\noindent
To do so, we look for possible dominant balances by substituting the forms $x(t)=\theta t^p,\quad y(t)=\xi t^q$ in the dominant part of the decomposition described by the system below
\begin{eqnarray}\label{ds1}
\dot{x}&=&y\nonumber\\
\dot{y}&=&-3(\alpha+1)xy.
\end{eqnarray}
 We assume here that   $\mathbf{\Xi}=(\theta, \xi)\in \mathbb{C}$ and  $\mathbf{p}=(p, q)\in \mathbb{Q}$. This leads to the unique balance
\begin{equation}
\mathcal{B}_{I}=[\mathbf{\Xi},\mathbf{ p}]=\left[\left(\frac{2}{3(\alpha+1)},-\frac{2}{3(\alpha+1)}\right),(-1,-2)\right],
\end{equation}\par\noindent
for $0<\alpha\leq1$. The subdominant part of the splitting (\ref{dc1}) satisfies
\begin{equation}
\frac{f^{(sub)}_{I}(\mathbf{\Xi},t^\mathbf{p})}{t^{\mathbf{p}-1}}=\left(0,-\frac{4\alpha}{3(\alpha+1)}\right),
\end{equation}\par\noindent\par\noindent
and is asymptotically subdominant \cite{p1}
 in the sense that
\begin{equation}\label{conditionofsubdomiannce}
\lim_{t\rightarrow 0}\frac{f^{(sub)}_{I}(\mathbf{\Xi},t^\mathbf{p})}{t^{\mathbf{p}-1}}\rightarrow 0,
\end{equation}\par\noindent
 only if $\alpha\rightarrow 0$. We therefore conclude that in the neighbourhood of the finite-time singularity the asymptotic solution is meaningful only in the limit of vanishing $\alpha$, that is in the absence of interaction between the two fluids.
 \par
 Next we calculate the Kovalevskaya matrix given by,
 \begin{equation}
\mathcal{K}_{I}=\mathcal{D}f_{I}(\mathbf{\Xi})-diag(\textbf{p}),
\end{equation}\par\noindent\par\noindent
 where $\mathcal{D}f$ is the Jacobian matrix of the decomposition. For
 this case the Kovalevskaya matrix reads
\begin{equation}     \mathcal{K}_{I}=\left[ {\begin{array}{cc}
   1 & 1 \\
   2 & 0 \\
 \end{array} } \right].
 \end{equation}\par\noindent
 
As discussed in \cite{met1}  the number of non-negative $\mathcal{K}$- exponents equals the number of  arbitrary constants expected to appear in the series solution, while the $-1$ exponent corresponds to the arbitrary constant relevant to the position of the singularity (for notational convenience taken to be $t=0$). Therefore, if the balance is to correspond to a general solution, two arbitrary constants are expected to appear in the series expansion (since the original system (\ref{ds}) is two dimensional). Here we find
 \begin{equation}
 spec(\mathcal{K}_{I})=(-1,2),
 \end{equation}\par\noindent
  with a corresponding eigenvector 
 \begin{equation}
  v_{2}^T=(1,-1).
 \end{equation}\par\noindent
 Hence, it is expected that  the balance $\mathcal{B}_{I}$ will correspond to a  general solution. Substituting the series expansions
 \begin{equation}\label{forms}
  x=\sum_{j=0}^{\infty}c_{j1}t^{j-1}, \quad y=\sum_{j=0}^{\infty}c_{j2}t^{j-2}, 
 \end{equation}\par\noindent\par\noindent
 in the system (\ref{ds}) we arrive after manipulations at the following asymptotic solution around the singularity
 \begin{equation}\label{sol1}
 x(t)=\frac{2}{3}t^{-1}+c_{21}t+\cdots,\quad as\quad t\rightarrow0,\quad \alpha\rightarrow0.
 \end{equation}\par\noindent
 The $y$-expansion is derived from the above by differentiation. 
 As a final test for the validity of this solution, a compatibility condition has to be satisfied for every positive $\mathcal{K}$-exponent \cite{p1}. For the positive eigenvalue  $2$ and an associated vector $ v_{2}^T=(1,-1)$ it reads
 \begin{equation}
 c_{21}=c_{22},
 \end{equation}\par\noindent
 and this is indeed true based on previous recursive calculations. 
\par\noindent 
 It follows from Eq. (\ref{sol1}) that all solutions are dominated by the $x=H\sim \frac{2}{3}t^{-1}$ solution which in terms of the scale factor reads
\begin{equation}\label{sf1}
a(t)\sim t^{2/3}\quad as\quad t\rightarrow0,\quad \alpha\rightarrow0.
\end{equation}\par\noindent
The dominant term of the series expansion (\ref{sf1}) is the same as the exact solution for the scale factor described by Eq. (\ref{scalef}) in the limit $\alpha\rightarrow0$. It also follows from Eq. (\ref{sf1}) that in the vicinity of the finite-time singularity and in the absence of interaction the total energy density of the model satisfies the form
\begin{equation}
\rho_{tot}\sim a^{-3}\quad t\rightarrow0,\quad\alpha\rightarrow0
\end{equation}\par\noindent
Now, the geometric character of the singularity is completely described in terms of the asymptotic behaviour of the total energy density and pressure of the model, the asymptotic behaviour of the scale factor and the Hubble parameter \cite{p2}. For the solutions above   it occurs that
\begin{equation}
\rho_{tot}\rightarrow \infty\quad P_{tot}\rightarrow0,\quad a\rightarrow\infty,\quad H\rightarrow\infty,
\end{equation}\par\noindent\par\noindent
as $t\rightarrow0$ and $\alpha\rightarrow0$.
The asymptotic conditions above describe the case of a Big Bang type of singularity.  Consequently, the singularity is necessarily placed at early times. It is discussed  in cf. \cite{Bento} that the contribution from the cosmological constant  is negligibly  small at early times, hence we conclude that our decomposition describes a model that  is indistinguishable from a CDM dominated universe in the past.
\par 
It is discussed in \cite{Bento} that the energy density  perturbations at early times regarding the dark matter component (and the baryon perturbations) are linear and small in scale ($\delta_m<<1$) and in the absence of interaction one can easily recover the standard energy perturbations in the CDM model. 
\section{\label{subsection2}Quasi de Sitter Universe}
\par\noindent
Let us now move on to the asymptotic analysis of the decomposition with dominant part given by the vector field
\begin{equation}\label{an}
f_{II}(x,y)=(y,-2\alpha y^2/x) . 
\end{equation}\par\noindent\par\noindent
Now by substituting in the asymptotic system $(\dot{x},\dot{y})=[y,-2\alpha (y^2/x)]$ the forms $x(t)=\theta t^p$ and $y(t)=\xi t^q$ we find the following dominant balance
\begin{equation}\label{bal2}
\mathcal{B}_{II}=\left[\left(\theta, \frac{\theta}{2\alpha+1}\right),\left(\frac{1}{2\alpha+1},\frac{1}{2\alpha+1}\right)\right],
\end{equation}\par\noindent
where $\theta$ is an arbitrary constant. The candidate subdominant part of the vector field, namely $f_{II}^{(sub)}(x,y)=[0,-3(\alpha+1)xy]$ is vanishing asymptotically without any restrictions on the values of the parameter $\alpha$ nor the constant $\theta$. Hence the decomposition is acceptable. To continue with, the Kovalevskaya matrix is given by 
\begin{equation}
\mathcal{K}_{II}=\left[ {\begin{array}{cc}
-1/(2\alpha+1) & 1 \\
(2\alpha)/(2\alpha+1)^2 & -(2\alpha)/(2\alpha+1) \\
\end{array} } \right],
\end{equation}\par\noindent
with corresponding eigenvalues
\begin{equation}
spec(\mathcal{K}_{II})={(-1,0)},
\end{equation}\par\noindent
and an  eigenvector 
\begin{equation}
v_{2}^T=\left(1,\frac{1}{2\alpha+1}\right).
\end{equation}\par\noindent
We note here that the second  $\mathcal{K}$-exponent is zero. Hence the arbitrary constants at the $j=0$ level of expansion (cf. \cite{p1,p2,met2,phd} for this terminology)
are the coefficients given by the dominant balance (\ref{bal2}), that is $(c_{01},c_{02})=\left(\theta,\frac{\theta}{2\alpha+1}\right)$. Therefore, the asymptotic solution is general since two arbitrary constants appear in the asymptotic solution as described below
\begin{equation}\label{sol2}
x(t)=\theta t^{1/(2\alpha+1)},\quad t\rightarrow0,
\end{equation}\par\noindent
for $0<\alpha\leq 1$. Since we are interested in expanding universes $(H>0)$, it follows that  the arbitrary constant $\theta$ attains only positive values. 
By integrating the solution above one obtains asymptotically the general solution for the scale factor described by the expression
\begin{equation}\label{sols}
a(t)=a_0\exp{\left(\theta C t^{1/C}\right)}\quad{as}\quad t\rightarrow0,
\end{equation}\par\noindent
where $C=(2\alpha+1)/(2\alpha+2)$.
\par
A comment about the asymptotic behaviour of the scale factor is in order. The specific form of Eq. (\ref{sols}) describes  an exponential evolution of the universe, with slower rate of expansion than the de Sitter universe, valid for a time interval. Clearly, as interaction kicks off the transfer of energy from dark matter to dark energy (described by Eqs. (\ref{c1})-(\ref{c2})) results in an important growth of the energy density of the vacuum. However the presence of dark matter decelerates the rate of  expansion. 
\par
As shown in the asymptotic solution (\ref{sols}) the $\alpha$ parameter determines the asymptotic states of the universe. For $0<\alpha\leq 1$ the universe enters (for a time interval) a quasi de Sitter space where the total energy density, total pressure, scale factor and Hubble parameter are asymptotically equal to 
\begin{equation}\label{forms}
\rho_{tot}\rightarrow0,\quad |P_{tot}|\rightarrow \infty,\quad a\rightarrow a_0,\quad H\rightarrow 0,
\end{equation}\par\noindent\par\noindent
respectively as $t\rightarrow0$, while  higher derivatives of $H$ diverge. This is a new type of singularity, a combination of   Type $IV$ \cite{sing,sd2,sd3,sd4} and Type $II$ (sudden) singularity placed at late times.
\par
It is interesting to note here that the present  decomposition describes an intermediate phase in the evolution of our interacting model. In the limiting case $\alpha\rightarrow 0$ (limit of no interaction) it is expected that the universe asymptotically (as $t\rightarrow0$) will approach the CDM model. This is indeed true since for $\alpha\rightarrow0$ the asymptotic analysis is identical to the one performed for the first decomposition in section \ref{subsection1}. Consequently, the particular decomposition  successfully reproduces the CDM model (as $t\rightarrow0$ and $\alpha\rightarrow0$).
\par
In addition,  it is also expected at late times  that the dominance of dark energy will drive the evolution of the universe towards de Sitter space. This is indeed feasible in the limit $\alpha\rightarrow\infty$, that is the case  where energy is being transferred  from dark matter to dark energy without bound. This results in the following asymptotic forms  for the total energy density, total pressure, scale factor and the Hubble parameter \footnote{If we assume here for purposes of notation that the  constant $\theta$ is positive and plays the role of the cosmological constant it can be proved that our solution (\ref{sols}) can also describe a  de-Sitter Universe  at a finite time at late epoch  $t_f\not=0$.}
\begin{equation}\label{sols1}
\rho_{tot}\rightarrow\rho_0,\quad |P_{tot}|\rightarrow\infty,\quad a\sim  a_0\exp{(\theta t)},\quad x=H\sim \theta,
\end{equation}\par\noindent\par\noindent
as $t\rightarrow0$, $\alpha\rightarrow\infty$. The forms above describe a dark energy dominated universe with a sudden type singularity placed at late times. It is discuss in \cite{Bento} that at late times the energy density perturbations of dark matter and baryons deviate from the linear behaviour  explaining the large energy transfer from dark matter to dark energy. 
\par
We note here that the exact solution described by Eq. (\ref{scalef}) fails to reproduce this specific behaviour of the scale factor at late times since it describes only possible power-law type solutions.

 \section{\label{subsection3}Interacting Vacuum}
\par\noindent
We now focus on the asymptotic analysis of all-terms-dominant case, that is the decomposition  (\ref{f3}), or equivalently described by the asymptotic system
\begin{eqnarray}\label{sys}
\dot{x}&=&y\nonumber\\
\dot{y}&=&-3(\alpha+1)xy-2\alpha\frac{y^2}{x}.
\end{eqnarray}\par\noindent
The subdominant vector field is the zero field in this case and there is one distinct balance given by
\begin{equation}\label{b3}
\mathcal{B}_{III}=\left[\left(\frac{2}{3},-\frac{2}{3}\right),(-1,-2)\right].
\end{equation}
\par\noindent
The Kovalevskaya matrix is given by 
\begin{equation}
\mathcal{K}_{III}=\left[ {\begin{array}{cc}
1 & 1 \\
4\alpha+2  & 2\alpha \\
\end{array} } \right],
\end{equation}\par\noindent
with corresponding eigenvalues
\begin{equation}
spec(\mathcal{K}_{III})={[-1,2(\alpha+1)]}.
\end{equation}
Even though the parameter $\alpha$ is present in the second $\mathcal{K}$-exponent, the form of the dominant  the balance (\ref{b3}) indicates that on approach to the finite-time singularity the dominant part of asymptotic solution $x\sim(2/3) t^{-1}$ is independent  from the choice of the parameter $\alpha$.\par
 For the whole series expansion though, the choice of the parameter $\alpha$ will determine the level of expansion at which the second arbitrary constant is expected to appear. For purposes of illustration we choose $\alpha=1/2$ so that $ spec(\mathcal{K}_{III})={(-1,3)}$. Then the associated eigenvector reads
\begin{equation}
 v_{2}^T=(1,-1).
\end{equation}\par
The candidate asymptotic solution is expected to be general if two arbitrary constants (the position of the singularity and one constant at the $j=3$ level of expansion) appear in the series expansion. After substituting the forms (\ref{forms}) into the asymptotic system (\ref{sys}), and for $\alpha=1/2$, we find the following asymptotic solution
    \begin{equation}\label{sol3}
        x(t)=\frac{2}{3}t^{-1}+c_{31}t^{-2}+\cdots,\quad t\rightarrow0.
    \end{equation}\noindent
    The compatibility condition at the $j=3$ level reads 
    \begin{equation}
        2c_{31}=c_{32}
    \end{equation}\par\noindent
    and it is indeed satisfied after recursive calculations. Hence the asymptotic solution found above is  general. In particular, the dominant behaviour of the solution (\ref{sol3}) on approach to the finite-time singularity is identical to the one of the decomposition $f_{I}(x,y)$. 
    \par
    We conclude here that the decomposition describes asymptotically  the model in the very early universe before interaction becomes significantly large. Having said that, the model described here has the same asymptotic features as in the case where the interaction is switched off asymptotically and the universe is matter dominated. Hence, at early times  the contribution of dark energy is negligible. 
\section{Discussion}\label{sec3}
\par\noindent
In this paper we analysed the stability of the singular flat space solutions that arise in the content of a unified dark energy model (GCG model) on approach to the finite-time singularity. We have shown that spacetime evolves from a phase that is initially dominated by dark matter to a phase that is asymptotically de-Sitter under some restrictions. The transition period in our model, between dark matter and dark energy domination corresponds to a quasi-inflationary regime that posses a new type of singularity asymptotically.\par
We conclude that the  current observational data are supportive towards an  asymptotically vanishing  interaction in a model where dark energy is simulated by a generalised Chaplygin gas cosmology. In particular, it is shown that for such unified model the interaction is  asymptotically vanishing at early times and the contribution of dark energy (as cosmological constant) is negligible. Hence,  the model is indistinguishable from CDM universe.  Such a model attains a pole-like \cite{phd} type of singularity and it is proved in previous works \cite{p1,p2} that such a dominant behaviour is an attractor of all possible asymptotic solutions on approach to the finite-time singularity.\par
An interesting era of expansion arises in an intermediate phase of expansion when the vector field decomposition  admits a quasi de-Sitter solution on approach to the finite-time singularity for $0<\alpha\leq 1$. In particular, the decomposition reproduces the successful CDM model at early times (in the limit as $\alpha\rightarrow0$) and approaches de-Sitter Universe at late times respectively. This intermediate phase of evolution is in alignment with the predictions of the GCG model for both early and late times. 
\par
To conclude with,  it would be interesting to apply the central projection technique of Poincar\`{e} to the dominant part of each of the asymptotic solutions on approach to the finite-time singularity to discuss the asymptotic stability of the model at infinity. This is examined in \cite{wip}.

\section{Acknowledgements}
\par\noindent
We thank Prof. David Wands and Prof. Elias C. Vagenas for discussions and useful comments.

\section{References}


\begin{thebibliography}{99}
\bibitem{mod}T.~Clifton, P.G.~Ferreira, A.~Padilla, C.~Skordis,  Physics Reports $\mathbf{513}$, 1-189 (2012) [arXiv: astro-ph/1106.2476].
\bibitem{obs1}C.L.~Bennett \emph{et al.} [WMAP Collaboration], Astrophys. J. Suppl. $\mathbf{208}$, 20 (2013), [arXiv: astro-ph/1303.5076].
\bibitem{obs2}A.G.~ Riess \emph{et al.} Astrophys. J. $\mathbf{730}$, 119 (2011), [arXiv: astro-ph/1103.2976].
\bibitem{c1}J.~Valiviita, E.~Majerotto, R.~Maartens, JCAP0807, 020 (2008), [arXiv: astro-ph/0804.0232].
\bibitem{c2}A.~Gromov, Y.~Paryshev and P.~Teerikorpi, Astron. Astrophys. $\mathbf{415}$, 813-820 (2004), [arXiv: astro-ph/0209458].
\bibitem{c3}G.~Caldera-Cabral, R.~Maartens and L.A.~Urena-Lopez, Phys. Rev. D $\mathbf{79}$, 063518 (2009), [arXiv: gr-qc/0812.1827].
\bibitem{c4}G.C.~CabraL, R.~Maartens and B.M.~Schaefer, JCAP 0907, 27 (2009), [arXiv: astro-ph/0905.0492].
\bibitem{co1}L.P.~Chimento, A.S.~Jakubi, D.~Pav\'{o}n and W.~Zimdahl, Phys. Rev. D $\mathbf{67}$, 083513 (2003), [arXiv: astro-ph/0303145]. 
\bibitem{co2}H.M.~Sadjadi and M.~Alimohammadi, Phys. Rev. D $\mathbf{74}$, 103007 (2006). 
\bibitem{Cha}W.~Zimdahl, D.~Pav\'{o}n and L.P.~Chimento, Phys. Lett. B $\mathbf{521}$, 133-138 (2001), [arXiv: astro-ph/0105479].
\bibitem{p1}S.~Cotsakis, G.~Kittou, Phys. Let. B $\mathbf{712}$, 16-21 (2012), [arXiv: gr-qc/1202.1407].
\bibitem{p2}S.~Cotsakis, G.~Kittou, Phys. Rev. D $\mathbf{88}$, 083514 (2013), [arXiv: gr-qc/1307.0377].
\bibitem{phd}G.E.~Kittou, Phd Thesis, University of the Aegean, 2015.
\bibitem{ba1}J.D.~Barrow and T.~Clifton, Phys. Rev. D $\mathbf{73}$, 103520 (2006).
\bibitem{ba2}T.~Clifton and J.D.~Barrow, Phys. Rev. D $\mathbf{73}$, 104022 (2006).
\bibitem{w1}D.~Wands, J.~De-Santiago and Y.~Wang, Class. Quant. Grav. \textbf{29 }, 145017 (2012), [arXiv:1203.6776]. 
\bibitem{Cha2}V.~Salvatelli, N.~Said, M.~Bruni, A.~Melchiorri and D.~Wands,  Phys. Rev. Lett. $\mathbf{113}$, 181301 (2014), [arXiv:  astro-ph/1406.7297].
\bibitem{Bento}M.C.~Bento, O.~Bertolami and A.A.~Sen, Phys. Rev. D $\mathbf{70}$, 083519 (2004), [arXiv: astro-ph/0407239].
\bibitem{met1}S.~Cotsakis and J.D.~Barrow, J. Phys. Conf. Ser. $\mathbf{68}$, 012004 (2007), [arXiv: gr-qc/0608137].
\bibitem{met2}A.~Coriely, Integrability and Nonintegrability of Dynamical Systems, World Scientific (2001).
\bibitem{2}A.Y.~Kamenshchik, U.~Moschella and V.~Pasquier, Phys. Lett.  B $\mathbf{511}$, 265-268 (2001), [arXiv:gr-qc/0103004].
\bibitem{Sand}H.B.~Sandvik, M.~Tegmark, M.~Zaldarriaga and I.~Waga, Phys. Rev. D $\mathbf{69}$, 123524 (2004), [arXiv: astro-ph/0212114].
\bibitem{Park}C.G.~Park, J.C.~Hwang, J.~Park and H.~Noh, Phys. Rev. D $\mathbf{81}$, 063532 (2010), [arXiv: astro-ph/0910.4202].
\bibitem{sing}S.~Nojiri, S.D.~Odintsov and S.~Tsujikawa, Phys. Rev. D $\mathbf{71}$, 063004 (2005), [arXiv: hep-th/0501025].
\bibitem{sd2}J.D.~Barrow, G.J.~Galloway and F.J.~Tipler, Mon. Not. Roy. Astron. Soc. $\mathbf{223}$, 835 (1986).
\bibitem{sd3}J.D.~Barrow, Class. Quantum Grav. $\mathbf{21}$, L79 (2004), 	[arXiv:gr-qc/0403084].
\bibitem{sd4}and Class. Quantum Grav. $\mathbf{21}$, 5619-5622 (2004).
\bibitem{wip}Work in progress.
\end{thebibliography}
\end{document}